\documentclass[12pt,reqno]{article}
\usepackage{amsmath,hyperref,amssymb}
\usepackage{subfigure,graphicx,url}

\usepackage{cite}
\newcommand{\bea}{\begin{eqnarray}}
\newcommand{\eea}{\end{eqnarray}}
\newcommand{\be}{\begin{equation}}
\newcommand{\ee}{\end{equation}}
\newcommand{\eeq}{\end{equation}}
\newcommand{\beq}{\begin{equation}}

\allowdisplaybreaks \numberwithin{equation}{section}

\DeclareSymbolFont{AMSa}{U}{msa}{m}{n}
\DeclareSymbolFont{AMSb}{U}{msb}{m}{n}
\DeclareMathSymbol{\fieldR}{\mathalpha}{AMSb}{"52}

\def\beq{\begin{equation}}
\def\eeq{\end{equation}}
\def\be{\begin{equation}}
\def\ee{\end{equation}}
\def\bea{\begin{eqnarray}}
\def\eea{\end{eqnarray}}

\newcommand{\ba}{\begin{eqnarray}}
\newcommand{\ea}{\end{eqnarray}}
\newcommand{\lp}{\left(}
\newcommand{\rp}{\right)}
\newcommand{\ls}{\left[}
\newcommand{\rs}{\right]}

\newcommand{\cZ}{{\cal Z}}
\newcommand{\nn}{\nonumber}

\begin{document}
\begin{flushright} \small
SU/ITP-13/09, EFI-13-12, NSF-KITP-13-113
\end{flushright}
\bigskip
\begin{center}
 {\large\bfseries  Mathieu Moonshine and ${\cal N}=2$ String Compactifications }\\[3mm]

 \

Miranda C. N. Cheng$^1$, Xi Dong$^{2,3}$, John F. R. Duncan$^4$, Jeffrey A. Harvey$^5$,\\
Shamit Kachru$^2$ and Timm Wrase$^{2,3}$

{\small\slshape
$^1$Universite Paris 7, UMR CNRS 7586, Paris, France \\
\medskip
$^2$ SITP, Department of Physics and\\
Theory Group, SLAC\\
Stanford University, Stanford, CA 94305, USA\\
\medskip
$^3$ Kavli Institute for Theoretical Physics\\ Santa Barbara, CA 93106, USA\\
\medskip
$^4$ Department of Mathematics\\
Case Western Reserve University, Cleveland, OH 44106, USA\\
\medskip
$^5$ Enrico Fermi Institute and Department of Physics\\
University of Chicago, Chicago, IL 60637, USA\\
\medskip
{\upshape\ttfamily chengm@math.jussieu.fr, xidong@stanford.edu, john.duncan@case.edu,
j-harvey@uchicago.edu, skachru@stanford.edu, timm.wrase@stanford.edu }
\\[3mm]}
\end{center}

\

\

\vspace{1mm}  \centerline{\bfseries Abstract}
\medskip

There is a `Mathieu moonshine' relating the elliptic genus of $K3$ to the sporadic group $M_{24}$.  Here, we give evidence that this moonshine extends to
part of the web of dualities connecting heterotic strings compactified on $K3 \times T^2$ to type IIA strings compactified
on Calabi-Yau threefolds.  We demonstrate that dimensions of $M_{24}$ representations govern the new
supersymmetric index of the heterotic compactifications, and appear in the Gromov--Witten
invariants of the dual Calabi-Yau threefolds, which are elliptic fibrations over the Hirzebruch surfaces
$\mathbb{F}_n$.

\bigskip
\newpage

\tableofcontents


\section{Introduction}

Monstrous moonshine \cite{ConwayNorton} is a mysterious relation between two very natural objects in mathematics: the largest of the sporadic simple finite groups (the Fischer-Griess Monster), and the simplest modular function (the J-function).  More generally, it relates conjugacy classes of the Monster to Hauptmoduls for genus zero subgroups of $SL(2,\mathbb{R})$.

This relationship is (partially) explained by the physics of a particular string compactification.  The chiral conformal field theory corresponding to a $\mathbb{Z}_2$ orbifold of the bosonic string on $\mathbb{R}^{24}/\Lambda$ where $\Lambda$ is the Leech lattice has as its partition function the J-function with constant term set equal to zero, and admits Monster symmetry. The relation between moonshine and vertex algebras of chiral conformal field theories is discussed in the mathematics literature in \cite{FLM, Borcherds}, and from an accessible physical point of view in \cite{DGH}.  A more general review of the story of moonshine appears in \cite{Gannon}.

It is encouraging that the Monstrous moonshine relations between fundamental objects from two a priori distinct areas of mathematics find a natural home in string theory.  However, it is fair to say that the string vacuum which appears here has not played a very central role in other developments in string theory and quantum gravity.  For this reason, moonshine has not yet had significant impact on our present understanding of string theory.

In 2010, Eguchi, Ooguri and Tachikawa (EOT) observed similar mysterious relations between the sporadic group $M_{24}$ and the elliptic genus of the (4,4) superconformal field theory with $K3$ target \cite{EOT}. This hints that generalizations of moonshine may be important in understanding more physically central string vacua.  $K3$ serves as the simplest non-trivial example of Calabi-Yau compactification, and has played a central role in duality relations between string theories \cite{Aspinwall}.  Further work has considerably elucidated, refined and generalized the EOT conjecture \cite{Miranda, Gaberdiel:2010ch, Gaberdiel:2010ca, Eguchi:2010fg, Gaberdiel, Gaberdiel:2012gf, Gaberdiel:2013nya, Taormina:2013jza, Taormina:2013mda, Umbral, Rademacher, Gannonproof}.  For fairly recent reviews, see \cite{Cheng-Duncan, Gaberdiel:2012um}.

In this note, we provide evidence that the Mathieu moonshine of \cite{EOT} extends to a richer structure visible also in 4d ${\cal N}=2$ string compactifications.  Such compactifications arise simply in two different ways: from heterotic strings on $K3 \times T^2$ (with suitable gauge bundles over the compactification manifold), or from type II strings on Calabi-Yau threefolds.
These two sets of compactifications are related by string duality \cite{Kachru-Vafa, FHSV}.
We show here that in the heterotic theories, with either arbitrary choices of gauge bundles and no Wilson lines or with all instantons embedded in one $E_8$ and only Wilson lines in the other $E_8$, the one-loop prepotential universally exhibits a structure encoding degeneracies of $M_{24}$ representations.  We demonstrate that this structure is also visible in the Gromov--Witten invariants of the dual type IIA Calabi-Yau compactifications.  Our results build on a large body of work on threshold corrections in heterotic strings \cite{Antoniadis,Louisone,Louistwo,Harvey-Moore,Kiritsis, Moore,Kawai,Lustone,Lusttwo,Stieberger} as well as recent advances in understanding the Gromov--Witten theory of elliptic Calabi-Yau manifolds \cite{Scheidegger, Klemm}.

\section{Universal threshold corrections in heterotic $K3$ compactifications}\label{sec:heterotic}

\subsection{Basic facts about ${\cal N}=2$ heterotic models}

A heterotic compactification on a manifold $X$ involves, in an obligatory way, data beyond the
sigma-model metric on $X$.  The heterotic string comes equipped with $E_8 \times E_8$ (or $Spin(32)/\mathbb{Z}_2$)
gauge fields in ten dimensions.  The Bianchi identity for the three-form field requires that one turns on
a non-trivial gauge bundle $V$ over $X$ whenever $c_2 (X) \neq 0$.  More precisely, in the simplest case without five-branes one requires:
\begin{equation}
\label{topcond}
c_2(X) = c_2(V_1) + c_2(V_2),~~c_1(V_{1,2}) = 0
\end{equation}
where $V_{1,2}$ are stable, holomorphic vector bundles embedded in the two $E_8$ factors.

We consider the case of 4d ${\cal N}=2$ compactifications on $K3 \times T^2$ with trivial bundles
on the $T^2$.
The internal worldsheet theory corresponding to the $K3$ compactification is then generically a conformal field theory with $(0,4)$ supersymmetry.
In this case, the first condition in (\ref{topcond}) implies
\begin{equation}
\label{kthreecond}
\int_X c_2 (V_1) + \int_X c_2 (V_2) = n_1 + n_2 = 24
\end{equation}
where we use the fact that $\int_{X} c_2(TX) = 24$ for $X$ a $K3$ surface, and $n_{1,2} \geq 0$ for a supersymmetric compactification.  Therefore, we must choose an embedding of 24 instantons into the $E_8 \times E_8$ gauge groups.

For the case $\int_X c_2(V_1) = 24$, a particular choice of bundle -- $V_1 = TX$ -- yields the
so-called `standard embedding'.  The gauge connection is set equal to the spin connection of the
$K3$ surface.  In this case, the $(0,4)$ worldsheet supersymmetry is enhanced to $(4,4)$.  Then
the elliptic genus of such a sigma-model will match the results computed in \cite{OY}, and
the observation of \cite{EOT} about the elliptic genus of such a model will continue to hold force.

This implies immediately that in a very different physical setting from type II compactification on $K3$ --
heterotic standard embeddings enjoy half the supersymmetry and completely different space-time physics -- one can expect $M_{24}$ to emerge. This is interesting in itself, but we will demonstrate now a considerably stronger result.

We will show that for each choice of $n_{1,2}$ consistent with (\ref{kthreecond}), there
is a universal physical quantity which is governed by dimensions of $M_{24}$ representations. However, while for the standard embedding we can conclude that there is Mathieu moonshine based on the enhanced worldsheet supersymmetry, this is a priori not clear for other cases. In \cite{toappear} (see also \cite{Harrison:2013bya}) we compute twining genera for simple $K3$ conformal field theories with a variety of different instanton embeddings and only (0,4) worldsheet supersymmetry. We successfully identify certain symmetries of the CFT with conjugacy classes of $M_{24}$ and therefore expect Mathieu moonshine for all instanton embeddings. Since the values of $n_{1,2}$ can only be varied by non-perturbative transitions in the heterotic string,
this is a much stronger result than any statement about the standard embedding.

\subsection{Threshold corrections and the new supersymmetric index}\label{sec:universality}

The new supersymmetric index \cite{Vafa} is defined by
\begin{equation}
\label{znew}
{\cal Z}_{new} = {1\over \eta(q)^2} {\rm Tr}_R J_0 e^{i\pi J_0} q^{L_0 - c/24} \bar{q}^{\overline{L_0} - \bar{c}/24}~,
\end{equation}
where ${\rm Tr}_R$ means that the trace is taken over the Ramond sector of the $(c,\bar{c}) = (22,9)$ internal CFT associated to the $K3 \times T^2$ factor plus the $E_8 \times E_8$ gauge bundle. The extra factor of ${1\over \eta^2}$ arises from the two extra 4d space-time bosons that are present in light-cone gauge string theory.

The new supersymmetric index is a natural object in 4d ${\cal N}=2$ heterotic compactifications because it counts the number of BPS states.  As discussed in \S3 of \cite{Harvey-Moore}, morally speaking,
\begin{equation} \label{eq:BPSstates}
{\cal Z}_{new} = -2i \left[ \sum_{\rm BPS~vectors} q^\Delta \bar q^{\bar \Delta} - \sum_{\rm BPS~hypers} q^{\Delta} \bar q^{\bar \Delta} \right]~.
\end{equation}
Therefore, any special properties or moonshine exhibited by ${\cal Z}_{new}$ reflect on important
properties of the BPS states in the compactified string theory.

Universal threshold corrections to gauge and gravitational couplings can be expressed in terms of this index \cite{Antoniadis}.   As described there and studied in greater depth in \cite{Harvey-Moore,Kiritsis,Moore}, one finds
\begin{equation}
\label{Deltais}
\Delta_{\rm gauge/grav} = \int {d^2\tau \over \tau_2} \left[ -{i \over \eta(q)^2} {\rm Tr}_R
\left( J_0 e^{i\pi J_0} q^{L_0 - c/24} \bar{q}^{\overline{L_0} - \bar{c}/24} F_{\rm gauge/grav}\right)
-b_{\rm gauge/grav} \right]~.
\end{equation}
Here,
\begin{equation}
F_{\rm gauge} = Q^2 - {1\over 8\pi \tau_2}\,, \qquad F_{\rm grav} = E_2(q) - {3\over \pi \tau_2} \equiv \hat{E}_2(q)\,,
\end{equation}
where $Q$ is some generator of a simple factor in the gauge group, and $E_2$ is the second Eisenstein series (see appendix \ref{sec:conventions} for its definition and our conventions). $b_{\rm gauge/grav}$ are the constant beta function coefficients that will not play an important role in the following.

Because the measure ${d^2\tau \over \tau_2^2}$ is modular invariant and the total integrand in equation \eqref{Deltais} must be modular invariant, it follows that
\begin{equation}
-\tau_2 {i \over \eta(q)^2} {\rm Tr}_R \left( J_0 e^{i\pi J_0} q^{L_0 - c/24} \bar q^{\overline{L_0} - {\bar c}/24} F_{\rm gauge/grav}\right)
\end{equation}
must be modular invariant.

Focusing on the gravitational threshold and pulling the factor of $F_{\rm grav}$ out of the trace
in (\ref{Deltais}), we find that $\tau_2 {\cal Z}_{new}$ should be a (non-holomorphic) modular form
of weight -2 (with a pole at the infinite cusp).  For the case of compactifications on $K3 \times T^2$ without Wilson
lines, the new supersymmetric index further factorizes as
\begin{equation}
\label{zkthree}
{\cal Z}_{new} = -2i {1 \over \eta(q)^2} {\Theta_{\Gamma_{2,2}} \over \eta(q)^2} {\cal G}_{K3}~.
\end{equation}
Here, we have defined the sum over windings and momenta on the torus
\ba
\Theta_{\Gamma_{2,2}}(q,\bar q;T,U,\bar T, \bar U) &=&  \sum_{p \in\, \Gamma_{2,2}} q^{\frac12 p_L^2} \bar{q}^{\frac12 p_R^2} =  \sum_{p \in\, \Gamma_{2,2}} q^{\frac12 (p_L^2-p_R^2)} e^{-2 \pi \tau_2 p_R^2} \nn\\
&=& \sum_{m_i, n_i \in \,\mathbb{Z}} e^{2\pi i \tau (m_1 n_1 + m_2 n_2) - {\pi \tau_2 \over T_2 U_2} \vert TU n_2 + Tn_1 - Um_1 + m_2\vert^2}~,
\ea
where $T$ and $U$ are the moduli that determine the metric and Kalb--Ramond field (or NSNS $B$-field) on the two torus. One can show (c.f. for example, equations (2.1.52) and (2.1.53) of \cite{Mirthesis}) that $\Theta_{\Gamma_{2,2}}$ is invariant under $\tau \to \tau + 1$ (since $\Gamma_{2,2}$ is an even lattice), and that $\tau_2 \Theta_{\Gamma_{2,2}}$ is modular invariant. The latter follows from Poisson resummation using the fact that $\Gamma_{2,2}$ is unimodular.
Therefore, we find that ${\cal G}_{K3}$ as defined in (\ref{zkthree}) should transform under
the modular group such that ${\cal G}_{K3} \over \eta(q)^4$ has weight -2.

One can now show by elementary reasoning that ${\cal G}_{K3}$, and hence ${\cal Z}_{new}$, is uniquely determined:

\noindent
$\bullet$
The definition of ${\cal Z}_{new}$ in (\ref{znew}) together with the existence of the bosonic string tachyon at $L_0=0$ and the fact that the left moving central charge is $c=22$ tells us that ${\cal Z}_{new}$ has a ${1 \over q}$ pole.

\noindent
$\bullet$
The same is true of
${{\cal G}_{K3} \over \eta(q)^{4}}$ , since $\Theta_{\Gamma_{2,2}}$ has only non-negative powers of $q$.

\noindent
$\bullet$
Removing the ${1\over q}$ pole by multiplying through by $\eta(q)^{24}$, we find that
$\eta(q)^{20} {\cal G}_{K3}$ must be a holomorphic modular form of weight 10.  Up to
multiplication by a constant factor, there is a unique such form, which is $E_4 E_6$. We recall the definition of the two Eisenstein series $E_4$ and $E_6$ in the appendix in equation \eqref{eq:EisensteinExpansion}.

We conclude that, independent of the choice of instanton numbers $n_{1,2}$, one has
\begin{equation}
\label{unique}
{\cal G}_{K3} = {\cal C} \,{E_4(q) E_6(q) \over \eta^{20}(q)}~,
\end{equation}
for some constant $\mathcal{C}$. This multiplicative constant ${\cal C}$ can be fixed to unity by various arguments.  For instance,
due to gravitational anomalies, 6d perturbative heterotic compactifications on $K3$ always
satisfy
\begin{equation}
\label{anomaly}
n_H - n_V = 244\,,
\end{equation}
where $n_H$ is the number of massless hyper-multiplets and $n_V$ is the number of massless vector-multiplets of the 6d ${\cal N}=1$ supersymmetry. (More generally, there is a term proportional to the number of massless tensor multiplets $n_T$, but this is determined to be $n_T = 1$ in perturbative heterotic models). The condition (\ref{anomaly}) fixes the coefficient of the $q^{1/6}$ term in the expansion (\ref{unique}), and in turn requires ${\cal C} \equiv 1$. Similar arguments about the universality of the new supersymmetric index can be found in \cite{Kiritsis, Moore, Lustone, Stieberger}.

\subsection{Discovering $M_{24}$ representations}\label{sec:M24}

For the particular case of the standard embedding \cite{Harvey-Moore} showed that the new supersymmetric index is related to the $K3$ elliptic genus (cf. appendix \ref{sec:conventions} for the definition of the $\theta_i(q,y)$)
\be
\cZ_{K3}^{elliptic}(q,y) = 8\ls \lp \frac{\theta_2(q,y)}{\theta_2(q,1)} \rp^2 +\lp \frac{\theta_3(q,y)}{\theta_3(q,1)} \rp^2 +\lp \frac{\theta_4(q,y)}{\theta_4(q,1)} \rp^2 \rs~.
\ee
For the case without Wilson lines one finds
\ba\label{eq:ZnewElliptic}
&&\hspace{-7mm}\cZ_{new} = \frac{i}{2} \frac{\Theta_{\Gamma_{2,2}}(q,\bar{q};T,U,\bar T, \bar U) E_4(q)}{\eta(q)^{12}} \ls \lp{\rm ch}_s^{SO(12)} + {\rm ch}_c^{SO(12)}\rp \cZ_{K3}^{elliptic}(q,-1) \right. \qquad \\
&&\hspace{3mm}+ \left. \lp {\rm ch}_b^{SO(12)} + {\rm ch}_v^{SO(12)} \rp q^{\frac14} \cZ_{K3}^{elliptic}(q,-q^{\frac12}) - \lp{\rm ch}_b^{SO(12)} -{\rm ch}_v^{SO(12)}\rp q^{\frac14} \cZ_{K3}^{elliptic}(q,q^{\frac12}) \rs \nn \,,
\ea
where the basic, vector, spinor and conjugate spinor characters of level 1 $SO(12)$ are given by
\ba
{\rm ch}_b^{SO(12)} &=& \frac12 \lp \lp\frac{\theta_3(q)}{\eta(q)} \rp^6 +\lp\frac{\theta_4(q)}{\eta(q)} \rp^6 \rp\,, \nn\\
{\rm ch}_v^{SO(12)} &=& \frac12 \lp \lp\frac{\theta_3(q)}{\eta(q)} \rp^6 -\lp\frac{\theta_4(q)}{\eta(q)} \rp^6 \rp \,, \nn \\
{\rm ch}_s^{SO(12)} &=& {\rm ch}_c^{SO(12)} =\frac12 \lp\frac{\theta_2(q)}{\eta(q)} \rp^6\,.
\ea
These characters arise for the standard embedding from the twelve free fermions. These fermions lead to a manifest $SO(12)$ gauge group that is enhanced to $E_7$. The extra factors of $q^{1/4}$ in the third and fourth term in \eqref{eq:ZnewElliptic} account for the fact that in the heterotic string we also have to sum over anti-periodic boundary conditions while the elliptic genus of $K3$ is defined by a trace over the Ramond sector only.

Following \cite{OY} we define ${\cal N}=4$ Virasoro characters
\begin{equation}
{\rm ch}_{h=1/4, l = 0}(q,y) =-\frac{i y^{\frac12}\theta_1(q,y)}{\eta(q)^3}  \sum_{n=-\infty}^\infty \frac{(-1)^n q^{\frac12 n(n+1)} y^n}{1-q^n y}\,,
\end{equation}
and
\begin{equation}
{\rm ch}_{h=n+1/4, l=1/2}(q,y) = q^{n-1/8}{\theta_1(q,y)^2 \over \eta(q)^3}~.
\end{equation}
These are the elliptic genera of the short and long representations of the ${\cal N}=4$ superconformal algebra.

Now one can expand $\cZ_{K3}^{elliptic}(q,y)$ as follows
\be
\cZ_{K3}^{elliptic}(q,y) = 24 \,{\rm ch}_{h=1/4, l = 0}(q,y) + \sum_{n=0}^\infty A_n {\rm ch}_{h=n+1/4, l=1/2}(q,y)~,
\ee
with
\begin{equation}\label{eq:Avalues}
{A_n} = {-2,90,462,1540,4554,11592,...}~.
\end{equation}
One recognizes the first few coefficients as simply related to the dimensions of irreducible representations of $M_{24}$ \cite{EOT}, while the higher coefficients can be decomposed in a way which is fixed by requiring the twining genera to behave appropriately \cite{Miranda, Gaberdiel:2010ch, Gaberdiel:2010ca, Eguchi:2010fg}.

Due to the relation between the new supersymmetric index and the elliptic genus of $K3$ \eqref{eq:ZnewElliptic} it is clear that the new supersymmetric index can be expanded in a similar way. Using our result from the last section we write
\begin{equation}\label{eq:Znew}
\cZ_{new} = -2 i \Theta_{\Gamma_{2,2}} {E_4(q) E_6(q) \over \eta(q)^{24}} \equiv \frac{i}{2} \Theta_{\Gamma_{2,2}} {E_4(q) \over \eta(q)^{12}} \times G(q)~.
\end{equation}
Then we can expand $G(q)=-4 E_6/\eta^{12}$ as
\begin{equation}
G(q) = 24 g_{h=1/4, l = 0}(q) + g_{h=1/4, l = 1/2}(q) \sum_{n=0}^{\infty} A_n q^n
\end{equation}
where
\ba
g_{h=1/4, l}(q) &=&  \lp{\rm ch}_s^{SO(12)} + {\rm ch}_c^{SO(12)}\rp {\rm ch}_{h=1/4, l}(q,-1) \nn \\
&+& q^{1/4}  \lp {\rm ch}_b^{SO(12)} + {\rm ch}_v^{SO(12)} \rp {\rm ch}_{h=1/4, l}(q,-q^{\tfrac12}) \nn\\
&-& q^{1/4} \lp{\rm ch}_b^{SO(12)} -{\rm ch}_v^{SO(12)}\rp {\rm ch}_{h=1/4, l}(q,q^{\tfrac12})\,.
\ea
Importantly, the $A_n$ take again the values given in \eqref{eq:Avalues}.

As we have argued in the previous subsection, in the absence of Wilson lines the new supersymmetric index takes always the form \eqref{eq:Znew} so that the above expansion is also possible for instanton embeddings for which the world sheet symmetry is $\mathcal{N}=(0,4)$. Thus we have found that for the case without Wilson lines and arbitrary instanton embeddings the new supersymmetric index and therefore the BPS states \eqref{eq:BPSstates} have a decomposition in terms of dimensions of irreducible representations of $M_{24}$.

\subsection{Compactifications with Wilson lines}
In this subsection we discuss the case where we embed all instantons in the first $E_8$ and keep track of the eight Wilson lines moduli $V^i$, $i=1, \ldots,8$ for the second $E_8$. For the standard embedding one finds \cite{Harvey-Moore} that the supersymmetric index takes the form
\ba\label{eq:ZnewWilson}
&&\hspace{-7mm}\cZ_{new} = \frac{i}{2} \frac{\Theta_{\Gamma_{10,2}}(q,\bar{q};T,U,V^i,\bar T, \bar U,\bar{V}^i)}{\eta(q)^{12}} \ls \lp{\rm ch}_s^{SO(12)} + {\rm ch}_c^{SO(12)}\rp \cZ_{K3}^{elliptic}(q,-1) \right. \\
&&\hspace{3mm} \left.+ \lp {\rm ch}_b^{SO(12)} + {\rm ch}_v^{SO(12)} \rp q^{\frac14} \cZ_{K3}^{elliptic}(q,-q^{\frac12}) - \lp{\rm ch}_b^{SO(12)} -{\rm ch}_v^{SO(12)}\rp q^{\frac14} \cZ_{K3}^{elliptic}(q,q^{\frac12}) \rs\,, \nn
\ea
with
\ba
&&\Theta_{\Gamma_{10,2}} = \sum_{p  \in\, \Gamma_{10,2}} q^{\frac12 p_L^2} \bar{q}^{\frac12 p_R^2}  =  \sum_{p \in\, \Gamma_{10,2}} q^{\frac12 (p_L^2-p_R^2)} e^{-2 \pi \tau_2 p_R^2}\\
&&= \sum_{\substack{m_j, n_j \in \,\mathbb{Z} \\ b_i \in\, \Gamma_{8,0}}} e^{2\pi i \tau \lp m_1 n_1 + m_2 n_2+\frac12 \sum_i b_i^2 \rp - {\pi \tau_2 \over T_2 U_2 -\frac12 \sum_i (V^i_2)^2} \left| (TU-\frac12 \sum_i (V^i)^2) n_2 + Tn_1 - U m_1 + m_2 + b_i V^i\right|^2}~,\nn
\ea
where $\Gamma_{8,0}$ denotes the $E_8$ root lattice. Note that $p_L^2-p_R^2$ is the length squared of the vector $(b_i;m_1,-n_1;n_2,m_2)$ with respect to the signature $(10,2)$ inner product and $p_R^2$ is proportional to the inner product of this vector with the moduli vector defined in (2.10) of \cite{Harvey-Moore}.

Since the $K3$ elliptic genus appears in \eqref{eq:ZnewWilson} exactly as in \eqref{eq:ZnewElliptic} in the previous subsection we can again expand the new supersymmetric index and find coefficients that are related to the dimensions of representations of $M_{24}$. Such an expansion is possible for arbitrary embeddings of all instantons in the first $E_8$ gauge group since we have only replaced $\Theta_{\Gamma_{2,2}} E_4$ with $\Theta_{\Gamma_{10,2}}$ and a universality arguments like the one presented in \S\ref{sec:universality} fixes $\cZ_{new}= -2i \Theta_{\Gamma_{10,2}} E_6/\eta^{24}$ uniquely. We can of course also turn off any number of Wilson lines. Thus we conclude that if we embed all instantons in the first $E_8$ and turn on an arbitrary number of Wilson lines in the second $E_8$, the new supersymmetric index and therefore the BPS states \eqref{eq:BPSstates} have a decomposition in terms of dimensions of irreducible representations of $M_{24}$.

\section{Type IIA compactifications on elliptic Calabi-Yau threefolds}\label{sec:IIA}

The heterotic string with $n_1 = 12+n, n_2 = 12-n$ instantons in the two $E_8$ factors is dual to a Calabi-Yau compactification on the elliptic
fibration over the Hirzebruch surface $\mathbb{F}_n$.  Special examples of this duality were first discovered in
\cite{Kachru-Vafa,FHSV}, and the duality has been better understood and mapped out in great detail in the F-theory setting in \cite{F-theory,F-theorytwo,BIKMSV}.

In the last section, we saw that the $q$-expansions of the ${\it integrands}$ in the one-loop threshold corrections for ${\cal N}=2$ compactifications of heterotic strings are controlled in part by dimensions of representations of $M_{24}$.  In this section, we will show that the same modular function which played
a starring role in \S\ref{sec:heterotic},
${E_4(q) E_6(q) \over \eta(q)^{24}},$
also appears universally in the Gromov--Witten invariants of the dual Calabi-Yau compactifications of type II strings.  That is, this modular
form governs certain terms in the 4d prepotential of these models.  Here, $q$ is the complexified K\"ahler class evaluated on the fiber of an elliptic Calabi-Yau, and the $q$-expansion is counting multi-wrappings of this fiber by worldsheet instantons correcting the prepotential.

It is not immediately obvious that this is related to the results of the previous section, since there one is discussing the integrand of a one-loop integral, and here one is discussing the space-time prepotential directly.
In \S\ref{sec:matching}, we will show that in fact the results of this section are implied by those of \S\ref{sec:heterotic} together with
string duality.

\subsection{Prepotentials of Calabi-Yau models}\label{sec:IIAprepot}

The prepotential obtained from type IIA compactifications on a Calabi-Yau threefold $M$ has the general form
\begin{equation}
\label{prepo}
F^{II} = -{1\over 6}\kappa_{ABC}t^A t^B t^C - {\chi(M) \zeta(3) \over 16\pi^3}
+ {1\over (2\pi)^3} \sum_{d_1,...,d_{h^{1,1}}} n_{d_1,...,d_{h^{1,1}}} Li_3\lp e^{-2\pi \sum_A d_A t^A}\rp~.
\end{equation}
Here, $\kappa_{ABC}$ are the triple intersection numbers of $M$, $\chi(M)$ is the Euler
character, and $n_{d_1,...,d_{h^{1,1}}}$ gives the instanton counting of genus 0 and multi-degree $d_A$. The polylogarithm $Li_k$ is defined as $Li_k(x) = \sum_{n=1}^\infty \frac{x^n}{n^k}$.

The Hirzebruch surface $\mathbb{F}_n$ is a $\mathbb{P}^1$ bundle over $\mathbb{P}^1$, and for this reason, the elliptic
Calabi-Yau over $\mathbb{F}_n$ always has at least three K\"ahler moduli.
Viewing $\mathbb{F}_n$ as a $\mathbb{P}^1$ bundle over $\mathbb{P}^1$, these are
 the moduli controlling the sizes of the two $\mathbb{P}^1$s $\,t_{B_{1}}$, $t_{B_{2}}$, and a fiber modulus $t_F$.  In the heterotic dual, these three
`universal' moduli correspond to the heterotic dilaton $S$, and to the $T$ and $U$ moduli parametrizing the K\"ahler and complex structure of the $T^2$.  Additional vector multiplet moduli exist at generic points in moduli space if $n$ is such that maximal Higgsing of either
$E_8$ is impossible; this happens for $n \geq 3$.  We will not be concerned with these additional
moduli in this section; activating them corresponds to turning on Wilson lines.

Therefore, we shall be interested in the Calabi-Yau threefolds $M$ dual to the maximally Higgsed heterotic models (with a given $n$) with
all Wilson lines turned off.  For these theories, the expansion of the prepotential (\ref{prepo}), as well as its higher-genus
analogues, can be further refined.  Let us write the free energy of the topological string with target $M$ as
\begin{equation}\label{eq:FgTop}
F(g_s,t^A) = \sum_{g=0}^{\infty} g_{s}^{2g-2} F^{(g)}(t^A)~.
\end{equation}
$F^{(0)}$ coincides with the ${\cal N}=2$ prepotential $F^{II}$. The $F^{(g)}$ for $g\geq 1$ control the couplings of the operators $(F^{grav}_+)^{2g-2} R_+^2$ \cite{Antoniadis:1993ze}, where the $+$ subscripts denote the self-dual parts, $F^{grav}$ is the field strength of the graviphoton and $R$ denotes the Riemann tensor.

In the duality with heterotic strings, the overall volume of the base $\mathbb{P}^1$, controlled by $t_{B_1}$, is dual to the heterotic dilaton S.  This follows from the results of \cite{AL}, if we view the space as a $K3$ fibration.  Hence, we will work in the limit:
\begin{equation}
{\rm Perturbative ~heterotic~limit:}~t_{B_1} \to \infty, ~q_{B_1} \equiv e^{-2\pi t_{B_1}} \to 0~.
\end{equation}
To make use of the beautiful results of \cite{Scheidegger,Klemm}, we do a double expansion of the $F^{(g)}(q^A)$ in the two base moduli, multiplying functions of the elliptic fiber modulus $q_F \equiv e^{-2\pi t_F}$,
\begin{equation}
F^{(g)}(q^A) = \sum f_{k,l}^{(g)}(q_F) \,q_{B_1}^k q_{B_2}^l~.
\end{equation}
The question of determining the prepotential and its higher-genus analogues now reduces to finding
the functions $f_{k,l}^{(g)}(q_F)$.  The terms which can be compared to perturbative heterotic computations are those with
$k=0$.

The recent papers \cite{Scheidegger,Klemm}, building on \cite{Yau, Alim:2007qj} and other earlier works,
found a set of recursion relations which determine the $f^{(g)}_{k,l}$.  Because they are functions of
$q_F$, it transpires that the $f^{(g)}_{k,l}$ are actually quasi-modular forms with weight depending on
$k,l$ as well as the integer $n$ parametrizing which Hirzebruch surface $\mathbb{F}_n$ appears as the base.

More precisely, the result of \cite{Scheidegger,Klemm} states
\begin{equation}
\label{fis}
f_{k,l}^{(g)}(q_F) = \left( {q_F^{1 \over 24} \over \eta(q_F)}\right)^{2p(k,l)} ~P_{2g-2+p(k,l)}(E_2(q_F), E_4(q_F), E_6(q_F)),
\end{equation}
where $P_{2g-2+p(k,l)}$ is a quasi-modular form of weight $2g-2 + p(k,l)$ and
\begin{equation}\label{eq:pkl}
p(k,l) = \frac{k}{2} \int_M c_2(M) \wedge J_2 + \frac{l}{2} \int_M c_2(M) \wedge J_1~.
\end{equation}
Here, $J_{1}$ and $J_2$ are the harmonic (1,1) forms appearing in the expansion of the K\"ahler form that control the real parts of $t_{B_1}$ and $t_{B_2}$, respectively. \footnote{Note that although $k$ counts the powers of $q_{B_1}$ it appears as coefficient of $\int_M c_2(M) \wedge J_2$ in \eqref{eq:pkl}.} Furthermore, the $f_{k,l}^{(g)}(q_F)$ satisfy the following recursion relation when $M$ is the fibration over $\mathbb{F}_n$ for $n=0,1,2$: \footnote{We thank E. Scheidegger for clarifying discussions about this recursion relation.}
\ba
\label{RR}
{\partial f^{(g)}_{k,l} \over \partial E_2} &=& {1\over 24} \sum_{h=0}^{g} \sum_{s=0}^k \sum_{t=0}^l
(n s(k-s) - s(l-t) - t(k-s)) f^{(g-h)}_{s,t} f^{(h)}_{k-s,l-t}\nn\\
&&- {1\over 24} (2kl +(n-2)k - 2l - n k^2) f^{(g-1)}_{k,l}\,,
\ea
where we take the $E_2$-derivative of functions $f^{(g)}_{k,l}$ as given in \eqref{fis} by differentiating the function $P_{2g-2+p(k,l)}(E_2,E_4,E_6)$ with respect to the first argument. To compare to the heterotic string results of \S\ref{sec:heterotic}, we want to focus on one-loop computations in the heterotic string-coupling expansion.  We should therefore study terms in the prepotential independent of $q_{B_1}$.

The simplest non-trivial term we can study is then $f^{(0)}_{0,1}(q_F)$.  The recursion relation
(\ref{RR}) tells us that
\begin{equation}
{\partial f^{(0)}_{0,1} \over \partial E_2} = 0~.
\end{equation}
Using the fact that $\int c_2(M) \wedge J_1 = 24$ (which follows from the discussion in e.g.
\cite{AL}), we find from (\ref{fis}) that $f^{(0)}_{0,1}$ should be fixed entirely by determining a
modular form of weight 10.  As mentioned before, such a form is unique up to multiplication by a constant and is given by $E_4(q_F) E_6(q_F)$.
Fixing the overall normalization by comparing with the known Gromov--Witten invariants one finds \cite{Scheidegger}
\begin{equation}
\label{fois}
f^{(0)}_{0,1}(q_F) = -{1\over 4\pi^3} {q_F E_4(q_F) E_6(q_F) \over \eta(q_F)^{24}}~.
\end{equation}
This in particular leads to a prepotential whose $q$-expansion in base moduli (multiplying appropriate quasi-modular forms of the fiber modulus) takes the form
\begin{equation}
\label{iiprepis}
F^{II} =  \ldots - {1\over 4\pi^3} {q_F E_4(q_F) E_6(q_F) \over \eta(q_F)^{24}} q_{B_2} + {\cal O}(q_{B_1},q_{B_2}^2) ~.
\end{equation}
This set of terms in the prepotential (in the $n=0$ model) also played an important role
in the discussion of \cite{Pandharipande}.

While the modular form ${E_4 E_6 \over \eta^{24}}$ has played a starring role both in \S\ref{sec:heterotic} and
here, the matching between the results of the two sections is far from obvious at this stage.  In one case
the expression appeared as an integrand in a loop amplitude, while in the other it is the final
result for a (set of terms in the) prepotential. After we connect the two results in \S\ref{sec:matching}, it becomes clear that the universal heterotic result derived in \S\ref{sec:universality} together with string duality implies that the result above holds for all $n$, if we set the ``extra" vector multiplet moduli that generically appear (as Wilson lines of the generic un-Higgsed gauge factor) for $n\geq3$ to zero. It would be interesting to check this explicitly on the type II side.

\subsection{The higher genera}
From the form of the recursion relation \eqref{RR}, we see that generically the $E_2$-derivative of the $f_{k,l}^{(g)}$ for $g \geq 1$ involves $f_{0,1}^{(0)}$.
In the next section we will match the heterotic and type II results and thus connect $f_{0,1}^{(0)}$ to $M_{24}$. Due to the recursion relation it is then tempting to conjecture that all $F^{(g)}$ for $g\geq1$ should to a certain degree know about $M_{24}$. This conjecture is bolstered by our previous observation that the BPS states on the heterotic string side are governed by $M_{24}$ and the fact that the duality between type II and the heterotic string maps BPS states to BPS states.

In the dual heterotic compactifications one can calculate the $f^{(g)}_{k,l}$ for $k=0$ and for all $g$ at one-loop \cite{Marino:1998pg}, so let us look at the particular case of $k=0$ for which the recursion relation \eqref{RR} simplifies significantly. In particular the functions $f_{0,1}^{(g)}$ satisfy
\be\label{eq:recursionf01}
\partial_{E_2} f_{0,1}^{(g)} =\frac{1}{12} f_{0,1}^{(g-1)} \,.
\ee
For $g=1$ this means that
\be
f_{0,1}^{(1)} \propto \frac{E_2 E_4 E_6 + c_1 E_4^3 + c_2 E_6^2}{\eta^{24}}\,.
\ee
For the case of the elliptic fibration over $\mathbb{F}_0$ \cite{Klemm:2004km} find that $c_1=c_2=0$ so that $f_{0,1}^{(1)} \propto E_2 E_4 E_6/\eta^{24}$. This particular quasimodular form has appeared in the integrand of $\Delta_{\rm grav}$ in \eqref{Deltais}, which as we explain in the next section is not a coincidence but follows from string duality.

By induction we find from \eqref{eq:recursionf01} for all $g$ that
\be\label{eq:higherg}
(\partial_{E_2})^g f_{0,1}^{(g)} \propto f_{0,1}^{(0)} \propto {E_4 E_6 \over \eta^{24}} \,.
\ee
In the next section we relate the heterotic and type II results and clarify the connection between $f_{0,1}^{(0)}$ and $M_{24}$. The above recursion relation \eqref{eq:higherg} then links $M_{24}$ to all $f_{0,1}^{(g)}$. In \S\ref{sec:gravthreshold} we review how one calculates the $f_{0,l}^{(g)}$ for all $g$ explicitly on the heterotic side \cite{Marino:1998pg} and reproduce \eqref{eq:higherg}.

\section{Matching the results}\label{sec:matching}

Here, we demonstrate that in fact \S\ref{sec:M24} and \S\ref{sec:IIAprepot} give results which match precisely, due to the happy appearance of the unique modular form of weight 10 in both computations.
We first outline the proof in the particular case of the standard embedding, and then explain why the result is unaltered for other instanton embeddings. Here we rely on the beautiful results of \cite{Lustone, Stieberger}.

\subsection{Simplest case}\label{sec:matchingsimple}

In the heterotic theory the prepotential in the $STU$ model is given by
\be
F^{het} = STU + f^{1-loop}(T,U) + {\cal O}(e^{-2\pi S}) \,.
\ee
One can obtain a second-order differential equation for the one-loop prepotential, depending on the new supersymmetric index, by equating two distinct formulas for the low-energy gauge couplings appearing in \cite{Louisone,Louistwo}. Following \cite{Harvey-Moore} let us consider the case of the standard embedding, which implies $n_1 = 24, n_2 = 0$ with the second $E_8$ unbroken.

The running of the second $E_8$ gauge coupling depends on, among other things, the value of the new supersymmetric index with the insertion $Q^2(E_8) - {1\over 8\pi \tau_2}$
\begin{equation}
{\cal Z}^{E_8}_{new} = {1 \over \eta^2} {\rm Tr}_R \left(J_0 e^{i\pi J_0}  q^{L_0 - c/24} \bar q^{\overline{L_0} - \bar c/24}
\left[ Q^2(E_8) - {1\over 8\pi \tau_2}\right]\right)~.
\end{equation}
The supersymmetric index with the $Q^2$ insertion is a descendent of the new supersymmetric index \eqref{znew} and can be obtained by an appropriate $q$-derivative. For an $E_8$ gauge group we have to take the derivative of the $E_8$ current algebra and find
\be\label{eq:ZnewE8}
{\cal Z}^{E_8}_{new} = -2i \frac{\Theta_{\Gamma_{2,2}}}{\eta^{24}} \lp \frac{q \partial_q E_4}{8}-\frac{E_4}{8 \pi \tau_2} \rp E_6 = -{i\over 12}\Theta_{\Gamma_{2,2}} { \hat{E}_2E_4 E_6 - E_6^2 \over \eta^{24}} ~.
\ee
The two formulae for the low-energy gauge coupling for the heterotic string compactified on $T^2 \times K3$ are the following:

\noindent
1)  A direct one-loop calculation in the perturbative string theory \cite{Louisone} yields for the effective coupling of the $E_8$ gauge group:
\begin{eqnarray}
{1\over g^2(p^2)} &=& {\rm Re}(S) + {1\over 8\pi T_1 U_1} {\rm Re}(f^{1-loop} - (T_1 \partial_T + U_1 \partial_U) f^{1-loop})\nonumber \\
&+& {b(E_8) \over 16\pi^2} {\rm log}\left({M_s^2 \over p^2}\right) + {1\over 16\pi^2}
\int {d^2\tau \over \tau_2} (-i{\cal Z}_{new}^{E_8} - b(E_8))~.
\end{eqnarray}
The first line includes the tree-level and one-loop shift in the usual definition of the string coupling in terms of the dilaton.  The second line is a sum of the contributions coming from a usual field theory beta function (with $b(E_8) = -60$) and the one-loop string threshold corrections from charged massive states. $p$ denotes an energy scale below the string scale $M_s$.

\noindent
2) Calculating the one-loop exact effective gauge coupling in a (string-based) effective quantum field theory with cutoff at the Planck scale \cite{Louisone,Louistwo} one finds
\be\label{eq:couplingEFT}
{1\over g^2(p^2)} = {\rm Re}\left[ h(S,T,U)\right] +{b(E_8) \over 16\pi^2} \left( {\rm log}\left( {M_p^2 \over p^2}\right) + K(S,T,U,\bar{S},\bar{T},\bar{U}) \right)\,.
\ee
Here the Wilsonian gauge coupling is given by (cf. equation (4.39) of \cite{Louistwo})
\be
h(S,T,U) = S- {1\over 8\pi}\partial_T\partial_U f^{1-loop}- \frac{1}{8\pi^2}{\rm log}[J(iT) - J(iU)] -\frac{b(E_8)}{4\pi^2} {\rm log}[\eta(iT)\eta(iU)]\,,
\ee
and $K$ is the tree-level K\"ahler potential
\be
K(S,T,U,\bar{S},\bar{T},\bar{U}) = -{\rm log}\ls\tfrac12 (S +\bar S)\rs -{\rm log} \ls \tfrac12(T +\bar T)(U +\bar U)\rs\,.
\ee
Using that $M_p^2 = M_s^2 \,{\rm Re}(S)$ we can rewrite the gauge coupling \eqref{eq:couplingEFT} as
\begin{eqnarray}
{1\over g^2(p^2)} &=& {\rm Re}\left[ S - {1\over 8\pi}\partial_T\partial_U f^{1-loop} - {1\over 8\pi^2} {\rm log}[J(iT) - J(iU)]\right] \nonumber\\
&+& {b(E_8) \over 16\pi^2} \left( {\rm log}\left( {M_s^2 \over p^2}\right) - {\rm log}[2T_1 U_1]-4\,{\rm Re} \ls {\rm log}(\eta(iT)\eta(iU))\rs\right)~.
\end{eqnarray}
For energies $p$ well below the string scale both couplings have to be equal which leads to a differential equation for the one-loop prepotential in terms of the new supersymmetric index (with the $Q^2$ insertion). We find
\begin{eqnarray}
&~&-{\rm Re}\left(\partial_T \partial_U f^{1-loop} + {1\over T_1 U_1} ( f^{1-loop} - (T_1 \partial_T
+ U_1 \partial_U) f^{1-loop} ) \right) \nonumber \\
&=& {1\over 2\pi} \int {d^2\tau \over \tau_2} (-i{\cal Z}^{E_8}_{new} - b(E_8)) + {1\over \pi}
{\rm Re}({\rm log}[J(iT) - J(iU)]) \nn \\
&&+ {b(E_8) \over 2\pi} \lp {\rm log}[2T_1 U_1] +4\,{\rm Re} \ls {\rm log}(\eta(iT)\eta(iU))\rs\rp\,.
\end{eqnarray}
The integral
\begin{equation}
\int {d^2\tau \over \tau_2} (-i{\cal Z}_{new}^{E_8} - b(E_8))
\end{equation}
is evaluated explicitly in appendix A of \cite{Harvey-Moore}.  One finds \cite{Harvey-Moore} that the one-loop prepotential
\begin{equation}
\label{hetpre}
f^{1-loop}(T,U) = p(U,T) - {1\over 4 \pi^3} \sum_{\substack{k>0, l \in \mathbb{Z} \\ k=0, l>0}} c(kl) Li_3 \lp e^{-2\pi (kT + lU)}\rp ~
\end{equation}
solves the differential equation, where $p(U,T)$ is a specific cubic polynomial that is not relevant for us.  Here, the coefficients $c(m)$ are defined by
\begin{equation}
\label{cdef}
{E_4 E_6 \over \eta^{24}} = \sum_{m \geq -1} c(m) q^m = {1 \over q} - 240 + \cdots\,, \quad \text{and} \quad c(m)=0 \quad \forall m <-1\,.
\end{equation}
Using the identification $q_{B_2} = e^{-2\pi (T-U)}$, $q_F = e^{-2\pi U}$, we can expand the heterotic prepotential\footnote{The base modulus $t_{B_1}$ is proportional to the dilaton $S$ but there is no canonical geometric way to identify the dilaton \cite{Klemm:2004km}. This does not matter to us since we work to zeroth order in $q_{B_1}$.}
\be\label{eq:heteroticexpansion}
F^{het} = \ldots - {1\over 4 \pi^3} q_F q_{B_2} \sum_{l \geq -1} c(l) q_F^l + {\cal O}(q_{B_1},q_{B_2}^2)~.
\ee
Comparing \eqref{iiprepis} with \eqref{eq:heteroticexpansion}, we see that the perturbative heterotic corrections as a function of the moduli $T$ and $U$
match perfectly with the type IIA worldsheet instanton sum over rational curves which don't wrap the $\mathbb{P}^1$ dual to the heterotic dilaton, at least in the leading order of $q_{B_2}$.

\subsection{Other values of $n$}\label{sec:HetPerpotGeneral}

The previous section shows that the $M_{24}$ multiplicities encoded in the integrand for the one-loop threshold corrections in the heterotic theory indeed map directly over to the $M_{24}$ multiplicities visible in counts of rational curves, for the case $n=12$.  A similar proof goes through for all other values of $n$. If we embed $n_1$ instantons in the first $E_8$ and $n_2$ instantons in the second $E_8$ so that the second $E_8$ is broken to some subgroup $G$, one finds for the new supersymmetric index with a $Q^2(G) - {1\over 8\pi \tau_2}$ insertion (cf. equation (2.4) in \cite{Stieberger})
\ba\label{eq:ZnewG}
{\cal Z}^{G}_{new} &=& -{1\over 12}\Theta_{\Gamma_{2,2}} \left( \frac{ (E_2 - {3\over \pi \tau_2}) E_4 E_6 - \frac{n_1}{24} E_6^2- \frac{n_2}{24} E_4^3}{\eta^{24}} \right)\nn\\
&=& -{1\over 12}\Theta_{\Gamma_{2,2}} \left( \frac{ (E_2 - {3\over \pi \tau_2}) E_4 E_6 -  E_6^2}{\eta^{24}} -72(24-n_1) \right)~.
\ea
In the second line we have used the identity $E_4^3-E_6^2 = 1728 \eta^{24}$. Thus we can think of the different instanton embeddings as only changing the constant term in the $q$-expansion of $\frac{E_6^2}{\eta^{24}}$ by $72(24-n_1)$. This also changes the beta-function coefficients $b(G) \rightarrow b(G) + 6(24-n_1)$. Going through the calculation of \cite{Harvey-Moore} on finds that these two changes cancel in the calculation of the one-loop prepotential which is therefore again given by \eqref{hetpre}.

Alternatively, we can also obtain the one-loop prepotential from the more general results of \cite{Lustone, Stieberger}.  These authors turn on a Wilson line
in an $SU(2)$ factor of the gauge group remaining ``unHiggsed'' in the $E_8$ with more instantons. The resulting theory, in addition to the $S,T,U$ moduli, will have a modulus $V$ parametrizing the value of the Wilson line.

The heterotic prepotential then takes the general form
\begin{equation}
F^{het}(S,T,U,V) = S(TU - V^2) + f^{1-loop}(T,U,V) + {\cal O}(e^{-2\pi S})~.
\end{equation}
The one-loop prepotential $f^{1-loop}$ is determined by solving the differential equation analogous to that discussed in \S\ref{sec:matchingsimple}.  It is given by:
\begin{equation}
\label{npre}
f^{1-loop}(T,U,V) =  p(T,U,V) - {1\over 4 \pi^3} \sum_{k,l,b} \tilde c\left(4 kl - b^2\right) Li_3\lp e^{-2\pi (kT + lU + bV)} \rp~,
\end{equation}
where $p(T,U,V)$ is a cubic polynomial that will not be important for us. The $n_1$, $n_2$ dependent $\tilde c\left(4 kl - b^2\right)$ are defined via:
\begin{equation}\label{eq:E41E6expansion}
\frac{n_1}{24} \frac{E_{4}(q,y) E_6(q)}{\eta(q)^{24}} + \frac{n_2}{24} \frac{E_4(q) E_{6}(q,y)}{\eta(q)^{24}} = \sum_{m,b} \tilde{c}(4m-b^2) q^m y^b~.
\end{equation}
One can recover the results without Wilson line, for the prepotential of the three-modulus model by setting $V=0$ and doing the sum over multi-wrappings of the appropriate curve in \eqref{npre}.  Performing the sum over $b$ yields
\begin{equation}
F^{het} = \cdots -{1\over 4 \pi^3} \sum_{k,l} c(kl) Li_3 \lp e^{-2\pi (kT + lU)} \rp, \quad c(kl) \equiv \sum_b \tilde{c}(4kl-b^2)~.
\end{equation}
Since the sum over $b$ amounts to setting $y=1$ in \eqref{eq:E41E6expansion} it follows that the coefficients $c(kl)$ defined in this way are independent of $n_1$, $n_2$ and agree precisely with those in \eqref{hetpre}, \eqref{cdef}.

Above we have argued that the heterotic string theory leads to the same prepotential for all $n$, if we set the Wilson lines that generically arise for $n \geq 3$ to zero. On the type II side this corresponds to taking a singular limit of the elliptic fibration over $\mathbb{F}_{n\geq3}$ by setting certain K\"ahler moduli to zero. While we expect from string duality that the resulting prepotentials match in this singular limit, it is nevertheless illuminating to study it in more detail. For instance, Harvey and Moore \cite{Harvey-Moore} work out the prepotential for $n=12$ and keep track of the eight Wilson lines that generically arise in the universal threshold corrections through $\cZ_{new}$ as given in \eqref{eq:ZnewWilson}. They find that the one-loop prepotential has the following form
\ba
f^{1-loop}(T,U,V^i) &=&  p(T,U,V^i) -\frac{\hat{c}(0) \zeta(3)}{8 \pi^3} \nn \\
&&- {1\over 4 \pi^3} \sum_{\substack{k>0,l \in \mathbb{Z}, b_i \in \Gamma_{8,0}\\ k=0,l>0, b_i \in \Gamma_{8,0}\\k=l=0,b_i >0}} \hat{c}\left(kl - \tfrac12 (\textstyle{\sum_i} b_i^2) \right) Li_3\lp e^{-2\pi (kT + lU + b_i V^i)} \rp\,,\quad
\ea
where $p(T,U,V^i)$ is a for us irrelevant cubic polynomial and this time we did not absorb the constant term proportional to $\zeta(3)$ in the polynomial. The coefficients $\hat{c}(m)$ are defined by
\begin{equation}
{E_6 \over \eta^{24}} = \sum_{m \geq -1} \hat{c}(m) q^m = {1 \over q} - 480 + \cdots\,, \quad \text{and} \quad \hat{c}(m)=0 \quad \forall m <-1\,.
\end{equation}
Comparing with the type II prepotential \eqref{prepo} we find that the Euler characteristic of the elliptic fibration over $\mathbb{F}_{12}$ is correctly given by $\chi = 2 \hat{c}_0 = -960$. If we now take the limit $V^i=0$ we find that the constant term proportional to $\zeta(3)$ gets shifted since
\be
\sum_{k=l=0,b_i >0} \hat{c}\left(- \tfrac12 (\textstyle{\sum_i} b_i^2) \right) Li_3(1) = 120 \zeta(3) \,,
\ee
where we used that $Li_3(1)=\zeta(3)$, $\hat{c}(-1)=1$ and that there are 120 positive roots in the $E_8$ lattice that satisfy $\sum_i b_i^2 =2$. The constant term in the one-loop prepotential thus changes from
\be
-\frac{\hat{c}(0) \zeta(3)}{8 \pi^3} \rightarrow -\frac{(\hat{c}(0)+240) \zeta(3)}{8 \pi^3} = -\frac{c(0) \zeta(3)}{8 \pi^3}\,,
\ee
where $c(0)=-240$ as defined in \eqref{cdef}. Similarly, since $E_4(q)$ is the theta function associated to the $E_8$ root lattice it follows that
\be
\sum_{\substack{k>0,l \in \mathbb{Z}, b_i \in \Gamma_{8,0}\\ k=0,l>0, b_i \in \Gamma_{8,0}}} \hat{c}\left(kl - \tfrac12 (\textstyle{\sum_i} b_i^2) \right) Li_3\lp e^{-2\pi (kT + lU)} \rp = \sum_{\substack{k>0,l \in \mathbb{Z}\\ k=0,l>0}}c(kl) Li_3\lp e^{-2\pi (kT + lU)} \rp\,.
\ee
This explicitly verifies that for $n=12$ setting all the Wilson lines to zero gives our universal answer.

\subsection{Gravitational threshold corrections}\label{sec:gravthreshold}

Another quantity that receives only one-loop corrections and has been calculated in heterotic perturbation theory is the gravitational coupling i.e. the coefficient of the term $R^2$ \cite{Antoniadis:1992sa}. As we discussed in \S\ref{sec:universality} the universal threshold corrections for this coupling are determined by \eqref{Deltais}
\ba
\Delta_{\rm grav} &=& \int {d^2\tau \over \tau_2} \left[ -{i \over \eta(q)^2} {\rm Tr}_R \left( J_0 e^{i\pi J_0} q^{L_0 - c/24} \bar{q}^{\overline{L_0} - \bar{c}/24} \hat{E}_2(q) \right)-b_{\rm grav} \right] \nn\\
&=& \int {d^2\tau \over \tau_2} \left[ -\frac{2 i \Theta_{\Gamma_{2,2}} \hat{E}_2 E_4 E_6}{\eta^{24}}-b_{\rm grav} \right]~.
\ea
Since the integrand in this case is just $\hat{E}_2(q) \cZ_{new}$, it follows from our discussion in \S\ref{sec:heterotic} that it receives contributions only from BPS states and is independent of the particular instanton embedding. Moreover, it might be relevant for the Mathieu moonshine.

This coupling is actually the first of an infinite series of perturbative heterotic terms which are dual
to the type II higher-genus topological amplitudes $F^{(g)}$ introduced in \eqref{eq:FgTop}.  The
weakly coupled heterotic string captures the $F^{(g)}$ in the limit when the base $\mathbb{P}^1$ of the dual Calabi-Yau threefold is very large.
As mentioned above, the $F^{(g)}$ enter the action in the schematic form $\int F^{(g)} (F^{grav}_+)^{2g-2} R_+^2$, where the $+$ subscripts denote the self-dual parts, $F^{grav}$ is the field strength of the graviphoton and $R$ denotes the Riemann tensor. On the heterotic side the $F^{(g)}$ are one-loop exact and have been calculated in \cite{Antoniadis:1995zn}. They also only receive contributions from BPS states \cite{Harvey-Moore}. For the symmetric embedding with $n=0$ and in the weak coupling limit ($S \rightarrow \infty$), one can write a generating function for the $F^{(g)}$ \cite{Marino:1998pg}
\ba\label{eq:hethigherg}
F(\lambda,T,U) &=& \sum_{g=1}^{\infty} \lambda^{2g} F^{(g)}(T,U)\nn\\
&=& \sum_{g=1}^{\infty} \frac{\lambda^{2g}}{2 \pi^2 (2 T_2 U_2)^{g-1}} \int_\mathcal{F} \frac{d^2 \tau}{\tau_2} \tau_2^{2(g-1)} \frac{E_4 E_6}{\eta^{24}} \mathcal{P}_{2g} \sum_{p \in \,\Gamma^{2,2}} p_R^{2(g-1)} q^{\frac12 p_L^2} \bar{q}^{\frac12 p_R^2} ~,\qquad
\ea
where $\mathcal{P}_{2g}$ is a function of $\hat{E}_2$, $E_4$ and $E_6$ of weight $2g$ that can be explicitly calculated following the prescription in \cite{Marino:1998pg}. However, its precise form is not important to us.

The $F^{(g)}(T,U)$ in \eqref{eq:hethigherg} can be explicitly calculated \cite{Marino:1998pg} and (the antiholomorphic parts) turn out to be polylogarithms whose coefficients are given by the Gromov--Witten invariants of the dual type II geometry and are related to the expansion coefficients of $\frac{E_4 E_6 \mathcal{P}_{2g}}{\eta^{24}}$. The appearance of polylogarithms is particularly natural on the type II (or M-theory) side when writing topological string amplitudes in terms of Gopakumar--Vafa invariants. On the heterotic side one finds \cite{Marino:1998pg} that one of the terms in $\mathcal{P}_{2g}$ is proportional to $(E_2)^g$. This matches nicely with the recursion relation \eqref{eq:higherg} we found in the type II dual theories. We also note that the integrands in \eqref{eq:hethigherg} all contain a factor $\frac{E_4 E_6}{\eta^{24}}$ that is connected to $M_{24}$. For the case of $g=1$ this follows from our discussion in section \S\ref{sec:heterotic}. For $g\geq 2$ the explicit calculation of an expression very similar to \eqref{eq:hethigherg} in \cite{Antoniadis:1995zn} shows that $\frac{E_4 E_6}{\eta^{24}}$ arises in exactly the same way as for $g=1$. This also follows from the observation that schematically the $(F^{grav}_+)^{2g-2}$ part for $g\geq2$ arises from taking appropriate derivatives of the $\Theta_{\Gamma_{2,2}}$ factor in $\cZ_{new} =-2 i\Theta_{\Gamma_{2,2}} E_4 E_6/\eta^{24}=-2 i\Theta_{\Gamma_{2,2}} {\cal G}_{K3}/\eta^{4}$ since the graviphoton arises from the $T^2$ compactification. Thus the ${\cal G}_{K3}$ part that is related to $M_{24}$ remains unaltered and we find that all $F^{(g)}$ are related to $M_{24}$. Although equation \eqref{eq:hethigherg} was derived for $n=0$, we expect that it should be also true for all other instanton embeddings based on the universality argument for ${\cal G}_{K3}$ given in \S\ref{sec:universality}.  Of course, this will be true only on the locus where the ``extra" vector multiplet moduli which are generically present in the $n \geq 3$ cases are turned off, as otherwise dependence on the additional Wilson-lines can complicate the issue.

\section{Discussion}\label{sec:discussion}

It was recently discovered that the elliptic genus of the (4,4) supersymmetric sigma model with $K3$ target exhibits a moonshine relation to the Mathieu group $M_{24}$.  This hints at a hidden symmetry governing the BPS states of string theories with 16 supercharges.  Here, we saw that in theories with just 8 supercharges, corresponding to heterotic (0,4) compactifications on $K3 \times T^2$, a similar phenomenon occurs.  The new supersymmetric index has a factor ${E_6(q) \over \eta(q)^{12}}$ whose expansion involving $\mathcal{N}=4$ characters and characters of $SO(12)$ suggests a hidden role for $M_{24}$ in these theories.  The type II duals of the heterotic models are Calabi-Yau compactifications; the $q$-expansion of the new supersymmetric index appears there (in a subtle way) in the Gromov--Witten invariants.  These again determine (part of) the BPS spectrum of Calabi-Yau compactifications.

Our results suggest several broad questions for future exploration.

\noindent
$\bullet$
In the case of the original Monstrous moonshine conjectures, the McKay-Thompson series provided considerable further evidence for a relationship between distinct mathematical objects.  Here, a similar role can be played by the twining genera, as demonstrated already in the (4,4) avatar of Mathieu moonshine \cite{Miranda, Gaberdiel, Gaberdiel:2010ch, Gaberdiel:2010ca, Eguchi:2010fg}.  It is therefore interesting to examine explicit (0,4) conformal field theories, find appropriate discrete symmetries, compute the relevant twining genera, and see if they coincide with the prediction of Mathieu moonshine, at least in some cases. In fact, we have already found that twining genera do show interesting properties in specific CFTs illustrating the cases $n=0,4,12$.  We shall report on computations of twining genera, and identification of symmetries of certain particularly simple (0,4) $K3$ conformal field theories with certain subgroups of $M_{24}$, in \cite{toappear} (see also \cite{Harrison:2013bya}).

\noindent
$\bullet$
Another natural thought in the study of Mathieu moonshine has been the following.  It is an old
result of Mukai that the most general symplectic automorphisms of $K3$ are subgroups of $M_{23}$
\cite{Mukai}.  Recent investigations have explored the very natural question: could (4,4) conformal
worldsheet theories with $(c,\tilde c)= (6,6)$ that correspond to a $K3$ compactification
have `stringy symmetries' that enlarge the $M_{23}$ of Mukai to $M_{24}$, and can large
subgroups of $M_{24}$ arise in such theories \cite{Gaberdiel}?  The answers are somewhat surprising:
not all discrete symmetries (which commute with the ${\cal N}=4$ worldsheet SUSY) fit into $M_{24}$;
no such (4,4) theory realizes $M_{24}$ as a symmetry; however, all groups could potentially be contained in the Conway group
Co$_1$.

The moduli space of (0,4) conformal field theories is much larger than that of their (4,4) counterparts.
It would be natural to try and classify discrete symmetries of such theories; to find the maximal
order symmetries realized at any points in moduli space; and to see if any role for the large sporadic
simple finite groups seems natural in the answer to these questions.  A large number of (0,4)
theories are given as orbifolds of $K3$ (see \cite{Stieberger} for a nice list), and even more should be
realizable as (0,4) Landau-Ginzburg orbifolds with various gauge bundles using gauged linear
sigma model techniques along the lines of \cite{DK}.

\noindent
$\bullet$
In \S\ref{sec:M24}, where we discuss how dimensions of $M_{24}$ representations can be seen in the $q$-expansion of the modular forms arising in the new supersymmetric index, we expanded in characters of the ${\cal N}=4$ algebra without further comment.  This is arguably natural in the context of Mathieu moonshine for (4,4) theories.  However, in the (0,4) context, the left-movers (whose excited states the $q$-expansion is counting) do not naturally play well with any ${\cal N}=4$ superalgebra.
An expansion in Virasoro characters (perhaps extended by a $U(1)$ current algebra) would seem more natural.  Why do ${\cal N}=4$ characters still seem to play a natural role in reading off multiplicities of $M_{24}$ representations, even for (0,4) conformal field theories?

\noindent
$\bullet$
We have found evidence for $M_{24}$ moonshine in heterotic models with instantons in one $E_8$ and Wilson lines in the other $E_8$ gauge group as well as arbitrary instanton configurations with all Wilson lines turned off. It would be interesting to check whether this is also true for models that have both Wilson lines and instantons in the same $E_8$. The general heterotic prepotential for the particular case of one Wilson line is discussed in \cite{Lustone, Stieberger}. \cite{Lustone} also compare this to the type IIA prepotential for certain $n$.

\noindent
$\bullet$
On the type II side we have argued that the Gromov--Witten invariants (or equivalently Gopakumar--Vafa invariants) for Calabi-Yau threefolds which are elliptic fibrations over the Hirzebruch surfaces $\mathbb{F}_n$ are governed by the dimensions of representations of $M_{24}$. Since these Calabi-Yau spaces are $K3$ fibered, one might wonder whether $M_{24}$ also plays a role in the Gromov--Witten invariants for other $K3$ fibered Calabi-Yau manifolds, at least when one restricts to curves in the fibre.

\noindent
$\bullet$
For heterotic orbifold compactification to four dimensions, it was shown in \cite{Dixon:1990pc} that moduli dependent one-loop corrections to the gauge couplings are only non-trivial, if the orbifold has a  sector that preserves 4d $\mathcal{N}=2$ supersymmetry. There is a large class of orbifold compactifications that lead to 4d $\mathcal{N}=1$ supersymmetric theories and have twisted sectors that preserve $\mathcal{N}=2$ supersymmetry.   We expect that the Mathieu group $M_{24}$ will also play a role in these ${\cal N}=1$ compactifications. We are going to study this further in \cite{toappeartwo}.

\noindent
$\bullet$
Why does a factor of the theta function of affine $E_8$ current algebra ${E_4(q) \over \eta(q)^8}$ naturally arise in the index computation in \S\ref{sec:M24}? At a technical level this is true because e.g. the index is independent of $n$, and for $n=12$ there is an unbroken $E_8$ symmetry.  But one would like a more physical understanding of why the $E_8$ appears in theories with other values of $n$.   A related observation (that the states contributing to the index seem to appear in $E_8$ multiplets despite the evident absence of unbroken $E_8$ symmetry) appeared in
the original elliptic genus computations of \cite{OY}.  Perhaps this affine $E_8$ can be related to the one appearing in BPS spectra of certain non-critical strings in the Calabi-Yau description \cite{VafaMayr}.

\noindent
$\bullet$
It is interesting that the function $E_6(q)/\eta(q)^{12}$ also appears in monstrous moonshine: as the twining function attached to a commuting pair of $2A$ elements in the monster. As such the coefficients of $E_6(q)/\eta(q)^{12}$ admit an interpretation as dimensions of virtual, projective representations of the twisted Chevalley group $^2E_6(2)$, and our arguments therefore indicate a possible role for $^2E_6(2)$ in the counting of BPS states in $\mathcal{N}=2$ string compactifications. The precise physical interpretation for $^2E_6(2)$ in this setting is as yet unclear, but we can observe that it must play a role significantly different from that of $M_{24}$, for the decompositions into degrees of $M_{24}$-modules presented here suggest commuting actions of affine $SO(12)$ and the $\mathcal{N}=4$ algebra---actions which cannot be preserved by $^2E_6(2)$---and the twining functions attached to $M_{24}$ are generally mock modular, while the twining functions attached to (a suitable cover of) $^2E_6(2)$ are principal moduli or vanish identically according to the generalized moonshine conjectures. Also, $^2E_6(2)$ does not admit $M_{24}$ as a subgroup, so we can rule out the possibility that the observations about $M_{24}$ made here are specializations of an interpretation for $^2E_6(2)$.

\noindent
$\bullet$
From \cite{Umbral, Cheng:2013wca} we have seen that the $M_{24}$ moonshine is but one instance of the more general phenomenon of ``umbral moonshine". Moreover, different instances of umbral moonshine have a very similar structure given by the underlying Niemeier lattices \cite{Cheng:2013wca}. Another natural direction is to clarify the relation between the other instances of umbral moonshine and string theory compactifications. It is possible that to capture these other cases, one has to consider the wider class of heterotic compactifications with more general Wilson lines turned on.

One can hope that further study of ${\cal N}=2$ string vacua may shed light on the phenomenon of moonshine, or even more optimistically, that moonshine may help lead us to a more abstract reformulation of string vacua.

\bigskip
\centerline{\bf{Acknowledgements}}
We are grateful to S. Harrison, S. Hohenegger, D. Israel, A. Klemm, V. Kumar, E. Scheidegger, S. Stieberger and D. Whalen for conversations about related subjects. J.H. acknowledges the support of NSF grant 1214409. S.K. would like to acknowledge UCSF and the Aspen winter conference on ``Inflationary theory and its confrontation with data in the Planck era" for hospitality during the very early stages of this work. S.K. and X.D. are supported by the U.S. National Science Foundation grant PHY-0756174, the Department of Energy under contract DE-AC02-76SF00515, and the John Templeton Foundation. T.W. is supported by a Research Fellowship (Grant number WR 166/1-1) of the German Research Foundation (DFG). T.W. and X.D. would also like to acknowledge the KITP program on ``Primordial Cosmology'' for hospitality during part of this work. This research was supported in part by the National Science Foundation under Grant No. NSF PHY11-25915.

\appendix

\section{Conventions}\label{sec:conventions}
We use the following conventions for the Jacobi $\theta_i(q,y)$ functions
\ba
\theta_1(q,y) &=& i \sum_{n=-\infty}^{\infty} (-1)^n q^{\frac{(n-\frac12)^2}{2}} y^{n-\frac12}\,,\\
\theta_2(q,y) &=& \sum_{n=-\infty}^{\infty} q^{\frac{(n-\frac12)^2}{2}} y^{n-\frac12}\,,\\
\theta_3(q,y) &=& \sum_{n=-\infty}^{\infty} q^{\frac{n^2}{2}} y^n\,,\\
\theta_4(q,y) &=& \sum_{n=-\infty}^{\infty} (-1)^n q^{\frac{n^2}{2}} y^n\,,
\ea
where $q=e^{2 \pi i \tau}$ and $y=e^{2 \pi i z}$. Whenever the $y$-dependence is not specified, we have set $y=1$, for example $\theta_i = \theta_i(q) = \theta_i(q,1)$ and likewise for the other functions defined below.

We also use the Dedekind $\eta(q)$ function
\be
\eta(q) = q^\frac{1}{24} \prod_{n=1}^\infty (1-q^n) \,,
\ee
and the Jacobi--Eisenstein series $E_4(q,y)$ and $E_6(q,y)$, of index $1$, defined by
\ba
E_4(q,y) &=& \frac12 \lp \theta_2(q,y)^2 \theta_2^6 + \theta_3(q,y)^2 \theta_3^6+\theta_4(q,y)^2 \theta_4^6\rp\,,\\
E_6(q,y) &=& -\frac12 \lp \theta_2(q,y)^2 \theta_2^6 (\theta_3^4 + \theta_4^4) + \theta_3(q,y)^2 \theta_3^6 (\theta_2^4-\theta_4^4)-\theta_4(q,y)^2 \theta_4^6(\theta_2^4+\theta_3^4)\rp\,. \nn
\ea
Lastly we define the quasimodular Eisenstein series
\begin{equation}
E_2(q) = 1 - 24 \sum_{n=1}^{\infty} {nq^n \over {1-q^n}} = 1 - 24 \sum_{n=1}^{\infty} \sigma_1(n) q^n~,
\end{equation}
and the non-holomorphic modular form $\hat{E}_2(q) = E_2(q)-3/(\pi \tau_2)$, where $\sigma_k(n) = \sum_{d \vert n} d^k$ is the sum of the $k$-th powers of the divisors of $n$.

The Eisenstein series $E_4(q) = E_4(q,1)$ and $E_6(q)=E_6(q,1)$ can likewise be expanded as
\ba
E_{4}(q) &=& 1 + 240 \sum_{n=1}^{\infty} \frac{n^3 q^n}{1-q^n}  = 1+ 240 \sum_{n=1}^{\infty} \sigma_3(n) q^n \,,\nn\\
E_6(q) &=& 1 - 504 \sum_{n=1}^{\infty} \frac{n^5 q^n}{1-q^n} = 1 - 504 \sum_{n=1}^{\infty} \sigma_5(n) q^n~.\label{eq:EisensteinExpansion}
\ea
Finally let us recall that a modular form of weight $k$ (with trivial multiplier system) satisfies
\be
f\lp \frac{ a\tau+b}{c\tau+d} \rp = (c \tau+d)^k f(\tau), \quad \text{for all } \left(
                                                                          \begin{array}{cc}
                                                                            a & b \\
                                                                            c & b \\
                                                                          \end{array}
                                                                        \right) \in SL(2,\mathbb{Z}) \,,
\ee
and the Dedekind $\eta$ function has weight $1/2$ (and a non-trivial multiplier system), while $\hat{E}_2$, $E_4$ and $E_6$ transform as modular forms with weight 2, 4 and 6, respectively.

\bibliographystyle{JHEP}
\bibliography{refs}

\end{document}